\documentclass[11pt]{article}
\usepackage[totalwidth=385pt,totalheight=580pt]{geometry}
\usepackage{amsmath,amssymb,epsf,graphics,graphicx}
\usepackage{psfrag}
\newcommand{\hepth}[1]{{[arXiv:hep-th/#1]}}

\newcommand{\physics}[1]{{[arXiv:physics/#1]}}

\newcommand{\quantph}[1]{{[arXiv:quant-ph/#1]}}

\newcommand\blank[1]{}
\newcommand\CC{{\mathbb C}}
\newcommand{\fract}[2]{{\textstyle\frac{#1}{#2}}}

\renewcommand{\hat}{\widehat}
\newcommand\eq{\begin{equation}}
\newcommand\en{\end{equation}}
\newcommand\bea{\begin{eqnarray}}
\newcommand\eea{\end{eqnarray}}

\newcommand\ba{\(\begin{array}}
\newcommand\ea{\end{array}\)}
\newcommand{\resection}[1]{\setcounter{equation}{0}\section{#1}}

\begin{document}
\begin{titlepage}
\vskip 0.5cm
\vskip .7cm
\begin{center}
{\Large{\bf ABCD and ODEs }}
\end{center}
\vskip 0.8cm \centerline{Patrick Dorey$^1$, Clare Dunning$^2$,
Davide Masoero$^3$, Junji Suzuki$^4$ and Roberto Tateo$^5$} \vskip
0.9cm \centerline{${}^1$\sl\small Dept.\ of Mathematical Sciences,
University of Durham,} \centerline{\sl\small  Durham DH1 3LE, United
Kingdom\,}
\vskip 0.3cm \centerline{${}^{2}$\sl\small IMSAS, University of
Kent, Canterbury, UK CT2 7NF, United Kingdom}
\vskip 0.3cm \centerline{${}^{3}$\sl\small SISSA, via Beirut 2-4,
34014 Trieste, Italy}
\vskip 0.3cm \centerline{${}^{4}$\sl\small Department of Physics,
Shizuoka University, Ohya 836, SURUGA, Shizuoka, Japan.}
\vskip 0.3cm \centerline{${}^{5}$\sl\small Dip.\ di Fisica Teorica
and INFN, Universit\`a di Torino,} \centerline{\sl\small Via P.\
Giuria 1, 10125 Torino, Italy}
\vskip 0.2cm \centerline{E-mails:}
\centerline{p.e.dorey@durham.ac.uk, t.c.dunning@kent.ac.uk,}
\centerline{ masoero@sissa.it, sjsuzuk@ipc.shizuoka.ac.jp,
tateo@to.infn.it}

\vskip 1.25cm
\begin{abstract}
\noindent
We outline a relationship between conformal field theories and
spectral problems of ordinary differential equations, and discuss
its generalisation to  models related to classical Lie algebras.

\end{abstract}

\end{titlepage}
\setcounter{footnote}{0}
%%%%%%%%%%%%%%%%%%%%%%%%%%%%%%%%%%%%%
\def\thefootnote{\fnsymbol{footnote}}
%%%%%%%%%%%%%%%%%%%%%%%%%%%%%%%%%%%%%%%%%%%%%%%%%%%%%%%%%%%%%%%%%%%%%
%%%  start of the paper  %%%%%%%%%%%%%%%%%%%%%%%%%%%%%%%%%%%%%%%%%%%%
%%%%%%%%%%%%%%%%%%%%%%%%%%%%%%%%%%%%%%%%%%%%%%%%%%%%%%%%%%%%%%%%%%%%%
%
\resection{Introduction}
The ODE/IM
correspondence~\cite{Dorey:1998pt,Bazhanov:1998wj,Suzuki:1999rj,Dorey:2007zx}
has established a link between two dimensional conformal field
theory (CFT) and generalised spectral problems in ordinary
differential and pseudo-differential equations. It is based on an
equivalence between transfer matrix
eigenvalues~\cite{Bazhanov:1994ft, Bazhanov:1996dr} and Baxter
$Q$-functions in integrable models (IMs), and spectral determinants
\cite{Sha,Voros} of ordinary differential equations (ODEs).

In statistical mechanics, the transfer matrix and  its largest
eigenvalue -- denoted by $T$ in the following --  are central
objects. For example,  consider  the six-vertex model defined on a
square lattice with $N$ columns and $N'$ rows; $T$ can be written in
terms of  an auxiliary entire function $Q$ through the so-called
Baxter $TQ$ relation.  Up to an overall constant, $Q$ is completely
determined by the  knowledge of the positions of its zeros, the
Bethe roots, which are  constrained by the Bethe ansatz equations
(BAE).  Subject to
  some qualitative information  on the positions of the Bethe roots,
   easily deduced   by studying systems with small size,
 the Bethe ansatz leads to a
unique set of ground-state roots. In the  $N' \rightarrow
\infty$ limit the free energy per site $f$ is simply related to $T$ by
\eq
f \sim -{ 1 \over N} \ln T~.
\en
In
\cite{Bazhanov:1994ft,Bazhanov:1996dr}, Bazhanov, Lukyanov and
Zamolodchikov showed how to adapt the  same techniques  directly to
the conformal field theory (CFT) limit of the six-vertex model.
In this setting, we
 consider the conformal field theory with Virasoro  central
charge $c=1$ corresponding to the continuum limit of the six-vertex
model, defined  on an infinitely-long strip with twisted boundary
conditions along  the finite size direction. The largest transfer
matrix  eigenvalue $T$ depends on three independent parameters: the
(rescaled) spectral parameter $\nu$, the anisotropy $\eta$ and the
twist $\phi$. Defining $E$, $M$, $l$, $\omega$, $\Omega$ through the
following relations
\eq
E=e^{2 \nu},~~\eta={\pi \over 2} {M \over M+1},~~\omega=e^{i {\pi
\over M+1}},
~~\Omega=\omega^{2M},~~\phi={ (2l+1) \pi \over 2M+2}
\en
the resulting  $TQ$ relation is
\eq
T(E,l,M)Q(E,l,M)=\omega^{-\fract{2l+1}{2}} Q(\Omega E,l,M)+
\omega^{\fract{2l+1}{2}} Q(\Omega^{-1}  E,l,M)~.
\label{TTQ}
\en
The Baxter function $Q$ for this largest eigenvalue
is fixed by demanding entirety of both $T$ and $Q$, and
reality, positivity and `extreme packing' for $l>-1/2$ of the set
$\{ E_i\}$ of zeros of $Q$. The BAE follow from the entirety of $T$
and $Q$ via
\eq
Q(E_i)=0 \Rightarrow T(E_i)Q(E_i)=0 \Rightarrow { Q(\Omega E_i)
\over Q(\Omega^{-1} E_i)}
=-\omega^{2l+1}~.
\label{BA}
\en
 Surprisingly, equations (\ref{TTQ}) and (\ref{BA})
also emerge from an apparently  unrelated context: the study of
particular spectral problems for the  following differential
equation
\eq
\left (\left(\frac{d}{dx} - \frac{l}{x} \right)\left(\frac{d}{dx} + \frac{l}{x}\right) - x^{2M} +E
\right)y(x,E,l)=0~,
\label{eq5}
\en
\noindent
with   $x$  and  $E$ possibly complex. To see the emergence of
(\ref{BA}) from  (\ref{eq5}), we start from the unique solution
$\psi(x,E,l)$ of (\ref{eq5})
on the punctured complex plane $x
\in \CC\setminus \{0\}$ which has the  asymptotic
\eq
\psi \sim   x^{-M/2}
\exp(-\fract{1}{M+1}x^{M+1})~,~~~(M>1)
\label{asyp}
\en
as $|x| \rightarrow \infty$ in any closed sector contained in the
sector $|\arg x\,|<\frac{3\pi}{2M+2}$. This solution is entire in
$E$ and $x$\,. {}From $\psi$ we introduce a family of solutions to
(\ref{eq5}) using the `Sibuya trick' (also known as `Symanzik
rescaling'):
\eq
\psi_k = \psi(\omega^{k}  x,\;\Omega^{k}
E,\; \;l)~.
\label{symanzik}
\en
In (\ref{symanzik}), $k$ takes  integer values; any
pair $\{\psi_k,\psi_{k{+}1} \}$ constitutes a  basis of solutions
to (\ref{eq5}). An alternative way to characterize
a solution to (\ref{eq5}) is through  its behaviour near the origin
$x=0$. The indicial equation is
\eq
(\lambda-1-l)(\lambda+l)=0~,
\en
and correspondingly we can define two (generally) independent solutions
\eq
\chi^{+}(x,E)=\chi(x,E,l) \sim x^{l+1} +O(x^{l+3})\,,
\label{eq9}
\en
and $\chi^{-}(x,E)=\chi(x,E,-l-1)$, which transform trivially under
Symanzik rescaling as
\eq
\chi^{+}_k=\chi^{+}(\omega^{k} x,\Omega^{k} E)= \omega^{(l+1)k} \chi^{+}(x,E)~.
\label{trs}
\en
The trick is now to rewrite $\chi^{+}_0=\chi^{+}( x,E)$
respectively in terms of the basis $\{\psi_0, \psi_1 \}$ and
$\{\psi_{-1}, \psi_0
\}$:
\bea
2 i\chi^{+}_0&{=}&\omega^{-l-\frac{1}{2}}
 Q( \Omega
E)\psi_0 -Q(E) \omega^{-\frac{1}{2}}  \psi_{1}
\label{r1} \\
\hspace{-15pt} 2i \chi^{+}_0=2i\omega^{l+1}\chi^{+}_{-1}&{=}&
\omega^{\frac{1}{2}}Q(E) \psi_{-1}-\omega^{l+\frac{1}{2}}Q( \Omega^{-1} E)\psi_0
  \quad
\label{r2}
\eea
where the coefficients has been fixed by consistency among
(\ref{r1}), (\ref{r2}) and (\ref{trs}) and
\eq
Q(E,l) =W[\psi_0,\chi^+_0]~.
\en
Here $W[f,g]=f \frac{dg}{dx}-g \frac{df}{dx}$ denotes the Wronskian
of $f$ and $g$. Taking the ratio (\ref{r1})/(\ref{r2}) evaluated  at
a zero $E{=}E_i$ of $Q$ leads immediately to  the  Bethe ansatz
equations (\ref{BA}) without the need to introduce the  $TQ$
relation, though in this case it can be done very easily (see, for
example the recent ODE/IM review  article \cite{Dorey:2007zx}).
Correspondingly, $\chi$ becomes subdominant at $x
\rightarrow \infty$ on the positive real axis:
$\chi(x,E_i,l) \propto \psi(x,E_i,l)$. The motivation of  dealing
with $\chi$, instead of $\psi$ (\ref{asyp}), is two-fold. Firstly,
$\chi$ can be obtained  by applying the   powerful and numerically
efficient iterative method proposed by Cheng many years
ago~\cite{cheng:1962} in the context of Regge pole theory, and applied
to spectral problems of this sort in \cite{DDTb}.
 To this end we introduce the linear operator $L$, defined through its
 formal action
\eq
L[x^p] = \frac{x^{p+2}}{
  (p+l)(p-l-1) }\,.
\en
So for any polynomial ${\cal P}(x)$ of  $x$\,,
\eq
\left(\frac{d}{dx} - \frac{l}{x} \right)\left(\frac{d}{dx} +
\frac{l}{x}\right) L[{\cal P}(x)]= {\cal P}(x)~,
\label{dna}
\en
and the basic differential equation (\ref{eq5}), with the boundary
conditions (\ref{eq9}) at the origin, is equivalent to
\eq
\chi(x,E,l) = x^{l+1} + L\left[ (x^{2M}-E)
\chi (x,E,l) \right]~.
\label{chia}
\en
Equation  (\ref{chia}) is solvable by iteration and it allows the
predictions of the ODE/IM correspondence to be checked with very
high precision.

The initial results of~\cite{Dorey:1998pt,Bazhanov:1998wj,Suzuki:1999rj}
connected conformal field theories associated with the Lie algebra
$A_1$ to (second-order)
ordinary differential equations. The generalisation to
$A_{n-1}$-models was  established in
\cite{Suzuki:1999hu, Dorey:2000ma} but it was only
recently~\cite{Dorey:2006an}  that the ODE/IM correspondence was
generalised to the remaining classical Lie algebras $B_n$, $C_n$ and
$D_n$.  Our attempts to derive generalised $TQ$ relations from the
proposed set of pseudo-differential equations were unsuccessful, but
a series of well-motivated conjectures led us directly to the BAE,
allowing us to establish the  relationship between BAE and
pseudo-differential equation parameters. Moreover, while the
numerics to calculate the analogs of the functions $\psi$ turned out
to be very costly in CPU time, the generalisation of Cheng's method
proved very efficient and allowed very high precision tests to be
performed. This is our  second main reason to deal with  solutions
defined through the behaviour about $x=0$, rather than $x=\infty$.

\resection{Bethe ansatz for classical Lie algebras}
\label{BAe}
For any classical  Lie algebra $\mathfrak{g}$, conformal field theory
Bethe ansatz equations  depending on a set of $rank(\mathfrak{g})$
twist parameters $
\gamma{=}\{ \gamma_a \}$  can be written in a compact form as
\eq
\prod_{ b=1}^{rank(\mathfrak{g})} \Omega^{B_{ab}\gamma_b}_{\phantom
a} \frac {Q^{(b)}_{B_{ab}}(E^{(a)}_{i},\gamma)}
{Q^{(b)}_{-B_{ab}}(E^{(a)}_{i},\gamma)}= -1\,,\qquad i=0,1,2,\dots~
\label{dall0}
\en
where
$
Q^{(a)}_k(E,\gamma)=Q^{(a)}(\Omega^k E,\gamma),
$
and the numbers $E^{(a)}_i$ are the (in general complex) zeros
of the functions $Q^{(a)}$.
 In (\ref{dall0})  the indices $a$ and $b$ label
the simple roots of the Lie algebra $\mathfrak{g}$, and
\eq
B_{ab}= { (\alpha_a, \alpha_b) \over  |\hbox{\rm long
roots}|^2}~,~~~a,b=1,2,\dots,rank(\mathfrak{g})
\label{cab}
\en
where the $\alpha$'s  are the simple roots of $\mathfrak{g}$. The
constant
 $\Omega=\exp \left(i {2\pi \over h^{\vee} \mu} \right)$ is  a pure
 phase, $\mu$ is a positive
real number and  $h^{\vee}$ is the dual Coxeter number.

It turns out that the Bethe ansatz roots generally split into
multiplets (strings) with approximately equal modulus $|E_i^{(a)}|$.
The ground state of the model corresponds to
a configuration of roots containing only multiplets with a common
dimension $ d_a=K/ B_{aa}$; the model-dependent integer $K$
corresponds to the degree of fusion~(see for
example \cite{Kulish:1981gi}).

\resection{ The pseudo-differential equations}
To describe the pseudo-differential equations corresponding to
the $A_{n-1}$, $B_n$, $C_n$ and $D_n$ simple Lie algebras we first
introduce some  notation. We need an $n^{\rm
th}$-order differential operator~\cite{Dorey:2000ma}
\eq
D_n({\bf g})=D(g_{n-1}-(n{-}1))\,D(g_{n-2}-(n{-}2))\,\dots\,
D(g_1-1)\,D(g_0)~,
\label{dfactdef}
\en
\eq
D(g)=\left(\frac{d}{dx}-\frac{g}{x}\right)~,
\en
depending on $n$ parameters
\eq
{\bf g} {=}\{g_{n-1}, \dots,g_1, g_0 \}~~~,~~{\bf g^{\dagger}} {=}\{
n-1-g_0, n-1-g_1, \dots, n-1  -g_{n-1} \}~. \label{conj}
\en
Also, we introduce an  inverse differential operator $(d/dx)^{-1}$,
generally defined through its formal action
\eq
\left( { d \over dx} \right)^{-1} x^{s}= {x^{s+1} \over s+1}~,
\label{def00}
\en
and we replace the simple `potential' $P(E,x)=(x^{2M}-E)$ of
(\ref{eq5}) with
\eq
P_K(E,x)= ( x^{h^{\vee} M/K}-E)^K~.
\label{pk}
\en

Using the notation of Appendix B in~\cite{Dorey:2006an} the
proposed pseudo-differential equations are reported below. \\

{\bf \large $A_{n-1}$  models:}\\

\nobreak
\noindent The $A_{n-1}$  ordinary differential equations are
\eq
D_n({\bf g^\dagger})\chi^{\dagger}_{n-1}(x,E)=
P_K(x,E)\chi^{\dagger}_{n-1}(x,E)~,
\label{sunnq}
\en
\goodbreak
\noindent
with  the constraint $\sum_{i=0}^{n-1}g_i{=}\frac{n(n{-}1)}{2}$~ and
 the ordering $g_i < g_j< n-1,~\forall \ i<j$~.  We  introduce
the alternative set of  parameters  $\gamma=\gamma({\bf g})=\{
\gamma_a({\bf g})
\}$
\eq
\gamma_a= {2 K \over h^\vee M} \left(\sum_{i=0}^{a-1} g_i - {a(h^{\vee}-1)
\over 2} \right)~.
\en
\\
The solution $\chi^{\dagger}_{n-1}(x,E)$ is  specified  by its
$x\sim 0$ behaviour
\eq
 \chi_{n-1}^{\dagger} \sim x^{n-1-g_0}+\hbox{subdominant terms,}~~~ (x \rightarrow
 0^{+})~.
\label{x=0}
\en
In general, this function grows  exponentially  as $x$ tends to
infinity on the positive real axis. In Appendix B of
\cite{Dorey:2006an}, it was shown that the coefficient in front of
the leading term, but for an irrelevant overall constant, is
precisely the function   $Q^{(1)}(E,\gamma)$ appearing in the Bethe
Ansatz, that is
\eq
\chi_{n-1}^\dagger \sim Q^{(1)}(E,\gamma({\bf g})) \;x^{(1-n){M \over 2}} e^{ {x^{M+1} \over M+1} }+\hbox{subdominant terms,}
~~~ (x \rightarrow \infty)~.
\en
Therefore, the set of Bethe ansatz roots
\eq
\{E_i^{(1)} \}
\leftrightarrow Q^{(1)}(E_i^{(1)},\gamma)=0
\en
 coincide with
the  discrete set of $E$ values  in (\ref{sunnq}) such that
\eq
\chi_{n-1}^\dagger \sim o \left(x^{(1-n){M \over 2}} e^{ {x^{M+1} \over M+1}
} \right)~,~~~ (x \rightarrow \infty)~.
\en
This condition is equivalent to the requirement of absolute
integrability of
\eq
\left ( x^{(n-1){M \over 2}} e^{ -{x^{M+1}
\over M+1} } \right) \chi_{n-1}^\dagger(x,E)
\en
on the interval $[0,\infty)$.
It is important to stress that the  boundary problem defined above
for the function $\chi_{n-1}^\dagger$ (\ref{x=0}) is in general
different from the one discussed in Sections 3 and 4 in
\cite{Dorey:2006an} involving  $\psi(x,E)$. The latter function is
instead a solution to
the adjoint equation of (\ref{sunnq}) and  characterised by
recessive behaviour at infinity. Surprisingly, the two problems are
spectrally  equivalent and   lead to identical sets of Bethe ansatz
roots.

\vspace{0.5cm}

{\bf \large $D_{n}$  models:}\\

\begin{figure}
\centering
\includegraphics[width=4.2cm]{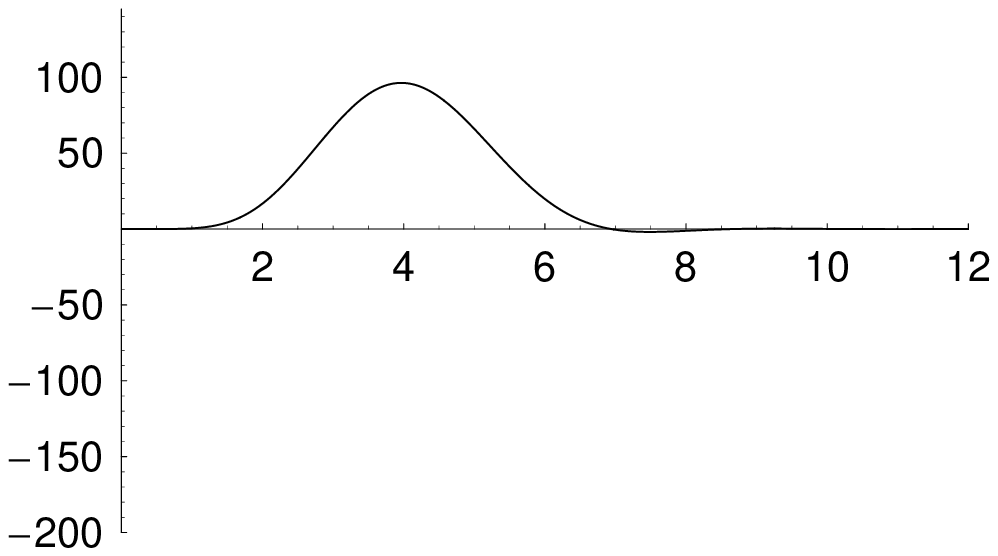}
\ \ \
  \includegraphics[width=4.2cm]{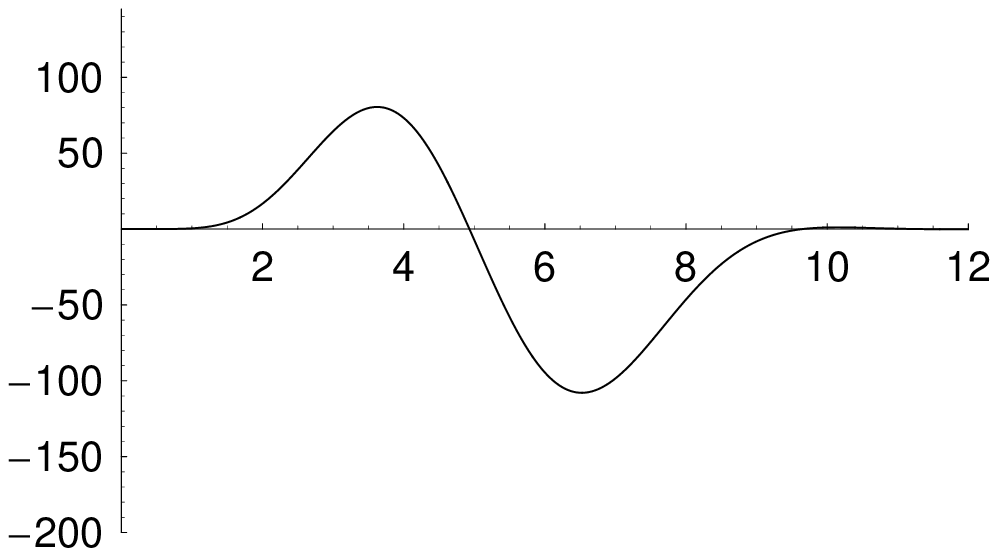}
\ \ \
  \includegraphics[width=4.2cm]{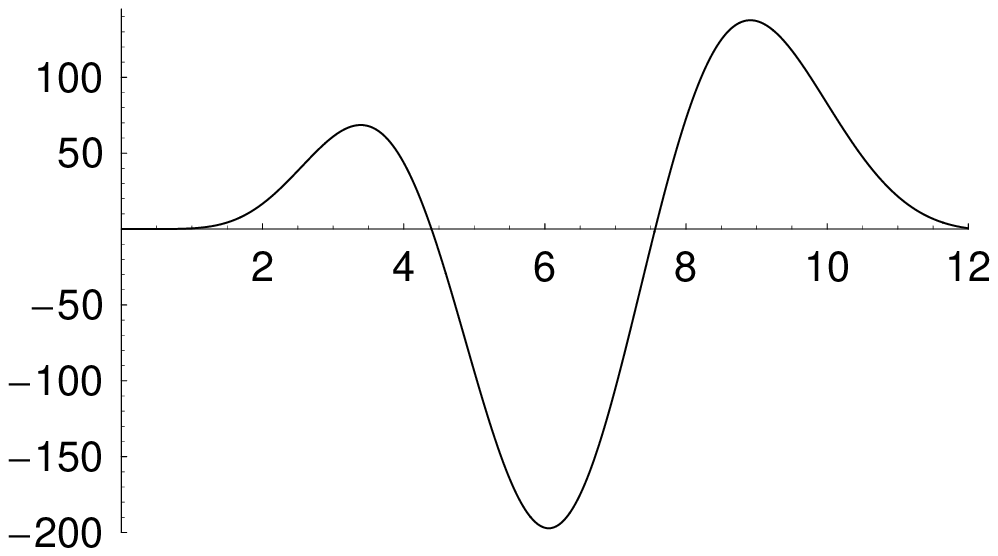}
    \caption{Lowest three functions $\Psi(x,E)$ for a $D_4$ pseudo-differential
      equation.\label{figwav}}
\end{figure}

\noindent The $D_{n}$  pseudo-differential equations are
\eq
D_{n}({\bf g^{\dagger}}) \left( \frac{d}{dx} \right)^{-1} D_{n}({\bf
g})  \chi_{2n-1}(x,E)=\sqrt{P_K (x,E)} \left(\frac{d}{dx} \right)
\sqrt{P_K (x,E)} \,\chi_{2n-1}(x,E)~.
\label{so2n02}
\en\\
Fixing   the ordering  $g_i<g_j<h^{\vee}/2$,   the ${\bf g}
\leftrightarrow \gamma$ relationship is
\eq
\gamma_a=  \frac{2K}{h^\vee M}\left( \sum_{i=0}^{a-1} g_i -
{a \over 2} h^{\vee} \right)~,~~(a=1,\dots, n-2)
\en
\eq
\gamma_{n-1}=
\frac{K}{h^\vee M}
\left(
\sum_{i=0}^{n-1} g_i - {n
\over 2}  h^{\vee}   \right),~\gamma_{n}= {K \over h^\vee M}  \left( \sum_{i=0}^{n-2} g_i -
   g_{n-1} - {n-2 \over 2}  h^{\vee}   \right)~.
\en
The solution is specified by requiring
\eq
\chi_{2n-1} \sim
x^{h^\vee-g_0}+\hbox{subdominant terms,}~~~(x \rightarrow  0^{+})~,
\en
\eq
\chi_{2n-1} \sim Q^{(1)}(E,\gamma({\bf g})) \;x^{-h^\vee{M \over 2}} e^{ {x^{M+1} \over M+1} }+\hbox{subdominant terms,}~~~(x \rightarrow \infty).
\en\\
Figure~\ref{figwav} illustrates $ \Psi(x,E)=x^{h^\vee {M \over 2}}
e^{ -{x^{M+1}
\over M+1} }  \chi_{2n-1}(x,E)$ for the first three
eigenvalues of the $D_4$ pseudo-differential equation defined by
$K{=}1,M=1/3$ and {\bf g}=(2.95,2.3,1.1,0.2)~.
\\

{\bf \large  $B_{n}$  models:}\\

\noindent The $B_{n}$ ODEs are
\eq
 D_{n}({\bf g^{\dagger}}) D_{n}({\bf g})  \chi^{\dagger}_{2n-1}(x,E) = \sqrt{P_K (x,E)}
\left(\frac{d}{dx} \right) \sqrt{P_K (x,E)} \chi^{\dagger}_{2n-1}(x,E)~.
\label{so2n102a}
\en\\
With the ordering $g_i<g_j<h^{\vee}/2$, the ${\bf g}
\leftrightarrow \gamma$ relation is
\eq
\gamma_a= {2 K \over h^\vee M}
\left(
\sum_{i=0}^{a-1} g_i - {a
\over 2}
h^{\vee} \right)~.
\en
The asymptotic behaviours about $x=0$ and $x=\infty$ are respectively
\eq
\chi_{2n-1}^{\dagger} \sim
x^{h^\vee-g_0}+\hbox{subdominant terms,}~~~(x \rightarrow  0^{+})~,
\en
and
\eq
\chi_{2n-1}^\dagger \sim Q^{(1)}(E,\gamma({\bf g})) \; x^{-h^\vee{M \over 2}} e^{ {x^{M+1} \over M+1} }+\hbox{subdominant terms,}
~~~(x \rightarrow \infty)~.
\en\\

{\bf  \large $C_{n}$  models:}\\

\noindent The pseudo-differential equations associated to the
$C_{n}$ systems are

\eq
 D_{n}({\bf g^{\dagger}}) \left(\frac{d}{dx} \right)D_{n}({\bf g})\,\chi_{2n+1}(x,E)
= P_{K }(x,E)  { \left(d \over dx \right)^{-1}}P_{K} (x,E)
\,\chi_{2n+1}(x,E)
\label{sp2n02}
\en\\
with the ordering  $g_i<g_j<n$.  The relation between the $g$'s  and
the twist parameters in the BAE is
\eq
\gamma_a= {2 K \over h^\vee M}
\left(\sum_{i=0}^{a-1} g_i - a n \right),~\gamma_n= { K \over h^\vee M}  \left( \sum_{i=0}^{n-1} g_i - n^2
\right)
\en
and
\eq
\chi_{2n+1}^{\dagger} \sim
x^{2n-g_0}+\hbox{subdominant terms,}~~~(x \rightarrow  0^{+})~,
\en
\eq
\chi_{2n+1}^\dagger \sim Q^{(1)}(E,\gamma) x^{-nM} e^{ {x^{M+1}
    \over M+1} }+\hbox{subdominant terms,}~~~ (x
\rightarrow \infty)~.
\en\\

Using a generalisation of Cheng's algorithm, the zeros of
$Q^{(1)}(E,\gamma)$ can be found numerically and shown to match
the appropriate Bethe ansatz roots~\cite{Dorey:2006an}.

\begin{figure}
\centering
\includegraphics[width=4.1cm]{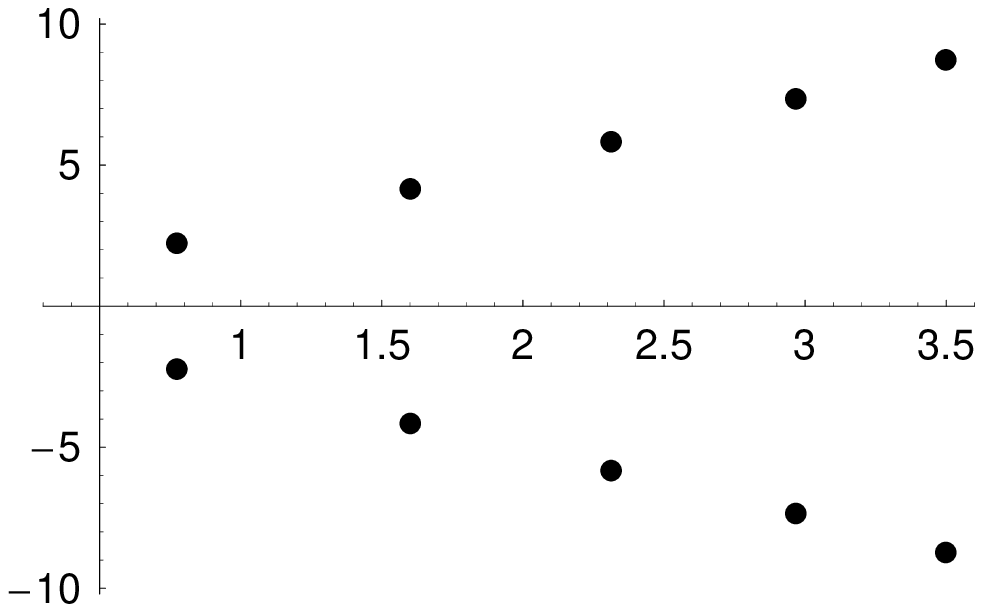}
\ \ \ \
  \includegraphics[width=4.1cm]{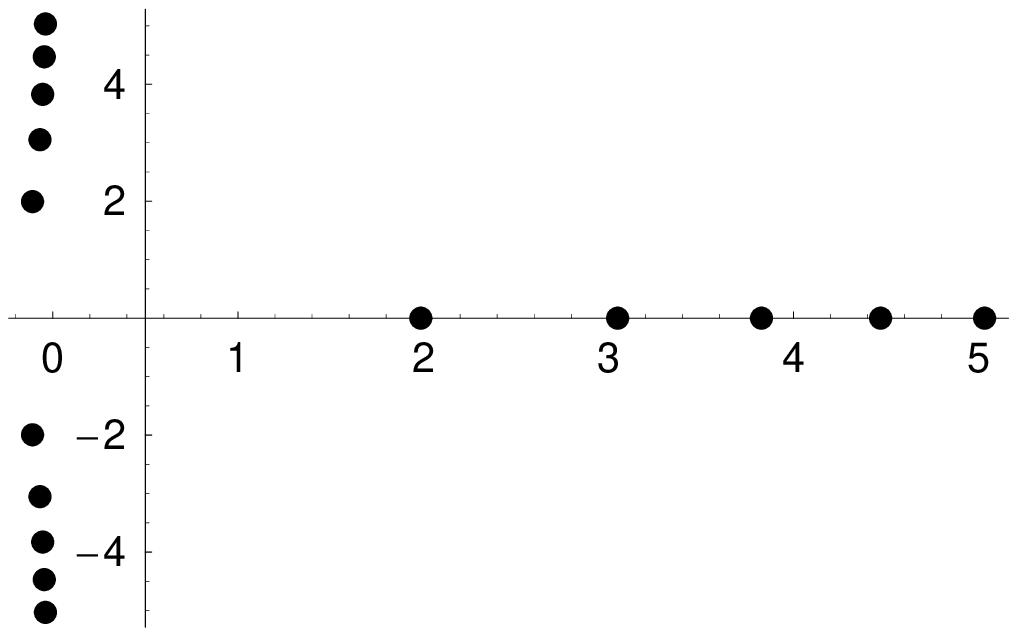}
\ \ \ \
  \includegraphics[width=4.1cm]{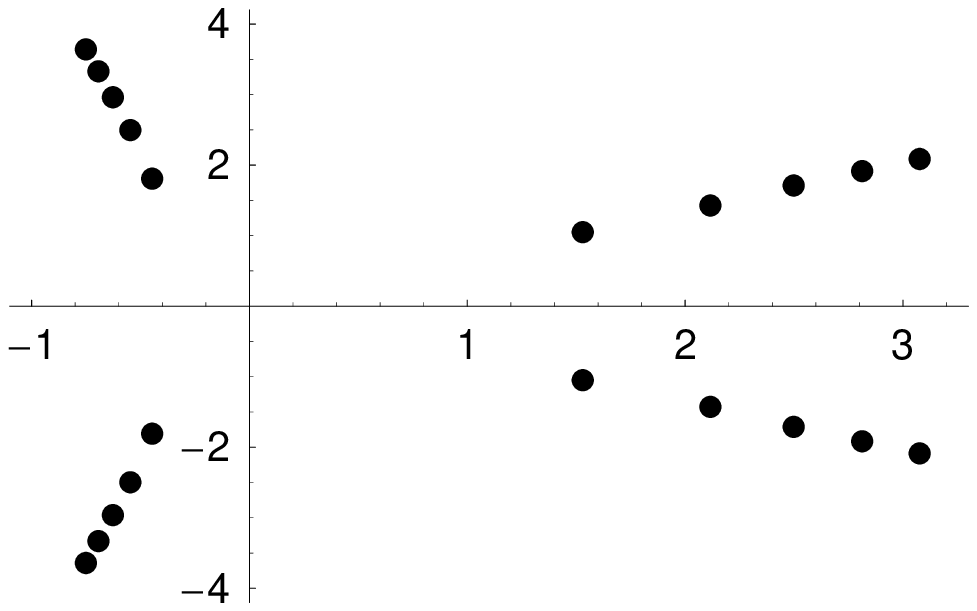}
    \caption{Complex $E$-plane: the eigenvalues for the $SU(2)$
model with $M=3$, $g_0=0$ for $K=2$, $3$ and $4$
respectively.\label{figpairs}}
\end{figure}

\begin{figure}
\centering
\includegraphics[width=4.1cm]{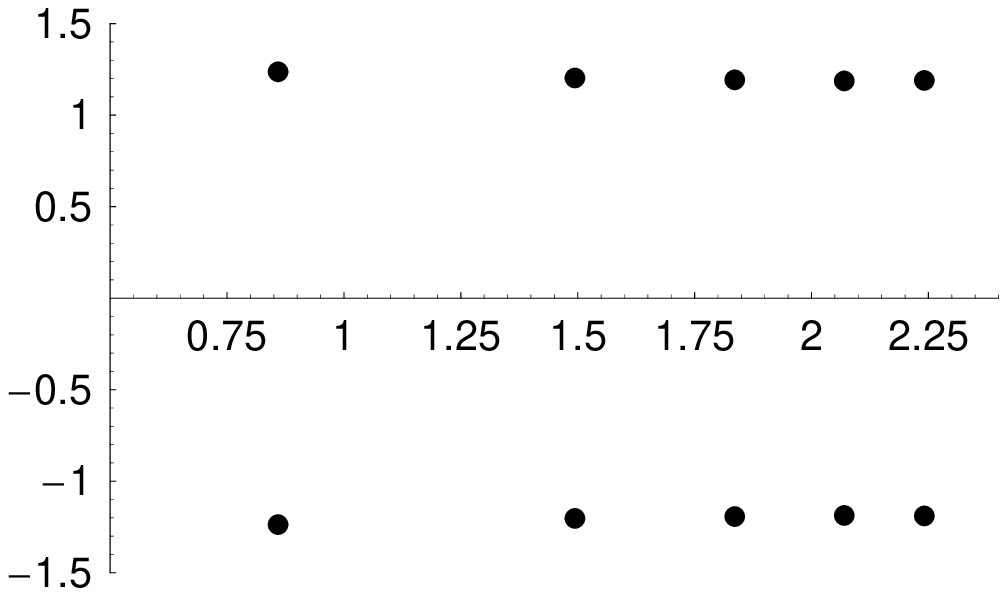}
\ \ \ \
  \includegraphics[width=4.1cm]{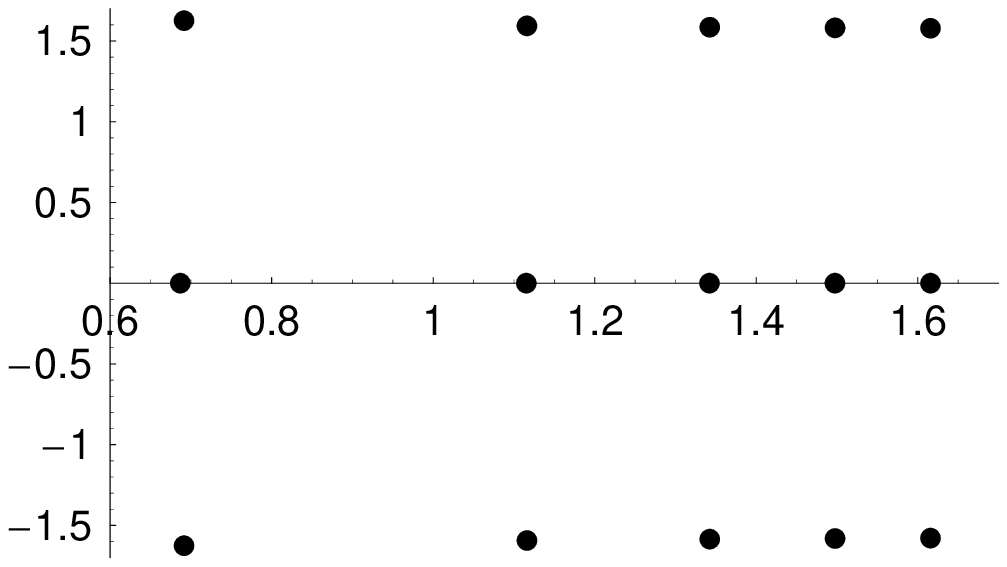}
\ \ \ \
  \includegraphics[width=4.1cm]{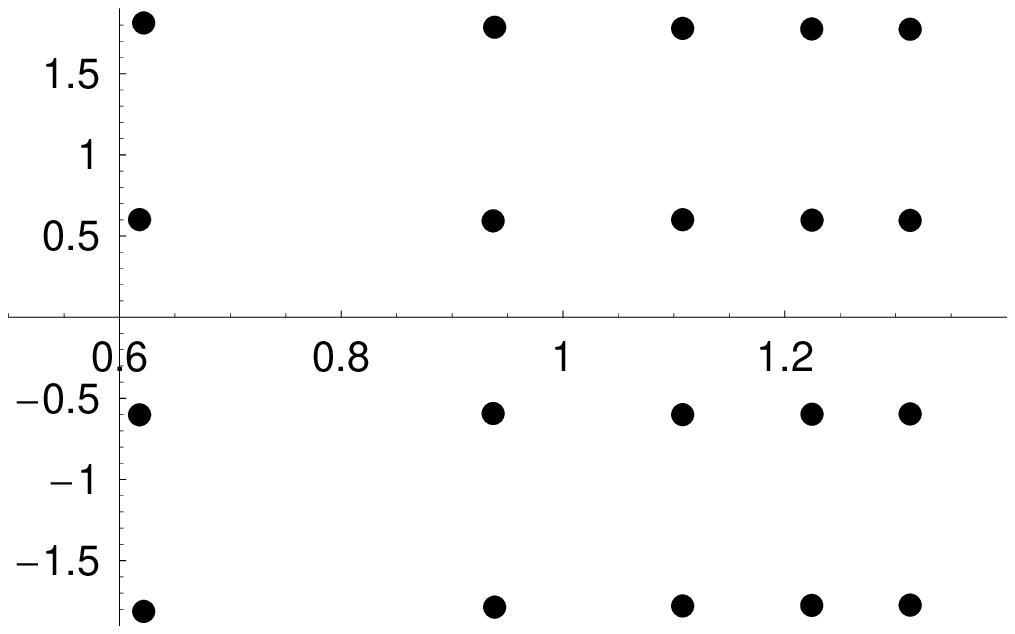}
    \caption{Complex $(\ln E)$-plane: two, three- and four-strings.
\label{twostrings}}
\end{figure}

In general, the   `spectrum' of a pseudo-differential equation may
be either real or complex.   In the  $A_{n-1}$, $B_n$, $D_n$ models
with $K{=}1$\footnote{ The $C_n$ spectrum is complex for any integer
$K{\ge}1$.}, the special choice $g_i=i$ leads to pseudo-differential
equations with real spectra, a property which is expected to hold
for a range of the parameters ${\bf g}$ (see, for example,
\cite{Dorey:2000ma}).
The $K{>}1$ generalisation of the potential (\ref{pk}), proposed
initially by Lukyanov for the $A_1$ models \cite{Luk-private} but
expected to work for all models,  introduces a new feature. The
eigenvalues corresponding to a $K{=}2,3$ and $K{=}4$ case of the
$SU(2)$ ODE are illustrated in figure~\ref{figpairs}.
The interesting feature appears if we instead plot the logarithm of
the eigenvalues as in figure~\ref{twostrings}.
We see that the logarithm of the eigenvalues form
`strings', a well-known feature of integrable
models.    The string solutions
approximately lie along lines in the complex plane, the deviations
away from which  can  be calculated
\cite{Dorey:2006an} using either
WKB techniques, or by studying the asymptotics of the Bethe ansatz
equations directly.

To end this section, we would like to comment briefly on  the
motivation behind  the conjectured pseudo-differential equations of
$B_n$, $C_n$ and $D_n$ type. Modulo the generalisation to $K{>}1$,
the $A_{n-1}$ type ODEs were derived in \cite{Dorey:2000ma}. We
began with the $D_3$ case since it coincides up to relabeling with
$A_3$, implying that the $D_3$ function $Q^{(1)}(E,\gamma)$ coincides
with the  $A_3$ function $Q^{(2)}(E,\gamma)$. Fortunately, the
latter is known~\cite{Dorey:2000ma} to encode the spectrum of a
differential equation satisfied by the Wronskian of two solutions of
the $Q^{(1)}$-related ODE.  The generalisation to $D_n$ models with
larger $n$ was then clear.  Further supporting evidence came from a
relationship between certain $D_n$ lattice models and the
sine-Gordon model, which appears as an $SU(2)$ problem.  This
relationship also extends to a set of $B_n$ models, and leads
naturally to the full $B_n$ proposal.    Finally, the $C_n$
   proposal arose from the $B_n$ cases via  a consideration of
   negative-dimension W-algebra
   dualities~\cite{Hornfeck:1994is}. Numerical and analytical  tests
   provided further
 evidence for the
  connection between these spectral problems and the Bethe ansatz equations
  for the   classical Lie algebras.

\section{Conclusions}
\label{conclusions}
The link between integrable models and the theory of ordinary
differential equations is  an exciting  mathematical fact that has
the potential to influence the future development of integrable
models and conformal field theory, as well as some branches of
classical and modern mathematics.  Perhaps the most surprising
aspect of the  functions $Q$ and $T$,  only briefly discussed in
this short note, is their variety of possible interpretations:
transfer matrix eigenvalues of integrable lattice models in their
CFT limit~\cite{Bazhanov:1994ft,Bazhanov:1996dr},  spectral
determinants of Hermitian and PT-symmetric~\cite{BB,BBN} spectral
problems~(see for example
\cite{DDTb}), g-functions of CFTs perturbed by relevant boundary
operators~\cite{Bazhanov:1994ft, Dorey:1999cj}, and particular
expectation values in the quantum problem of a Brownian
particle~\cite{Bazhanov:1998za}. Further, the (adjoint of the)
operators (\ref{sunnq}), (\ref{so2n02}), (\ref{so2n102a}) and
(\ref{sp2n02})
 resemble in form the Miura-transformed
Lax operators introduced by Drinfel'd and Sokolov in the context of
 generalised KdV equations, studied more recently in relation to the
 geometric Langlands
correspondence~\cite{Mukhin:2002fp2, Frenkel:2005fr}. Clarifying
this connection is an interesting open task. Here we finally observe
that the proposed equations respect the well-known Lie algebras
relations $D_2
\sim A_1
\oplus A_1$, $A_3
\sim D_3$, $B_1 \sim A_1$, $B_2 \sim C_2$. Also,  at special values of
the parameters the
 $C_n$ equations  are formally related  to the  $D_n$
ones by the analytic continuation $n
\rightarrow -n$, matching an interesting W-algebra duality
discussed by Hornfeck in~\cite{Hornfeck:1994is}:
\eq
{(\hat{D}_{-n})_K \times (\hat{D}_{-n})_L \over
 (\hat{D}_{-n})_{K+L}} \sim { (\hat{C}_{n})_{-K/2} \times
  (\hat{C}_{n})_{-L/2}
 \over  (\hat{C}_{n})_{-K/2 -L/2}}\,.
\label{dualdc}
\en
The relationship between our equations and coset conformal field
theories is another aspect worth investigation. We shall return to
this point in a forthcoming publication.

\medskip
\noindent{\bf Acknowledgements --}
RT thank  Vladas Sidoravicius, Fedor Smirnov and all the organizers
of the conference $M{\cap}\Phi$-- ICMP 2006 in Rio de Janeiro for
the invitation to talk at the conference and for the  kind
hospitality. JS thanks the Ministry of Education of Japan for a
`Grant-in-aid for Scientific Research', grant number 17540354. This
project was also partially supported by the European network EUCLID
(HPRN-CT-2002-00325), INFN grant TO12, NATO grant number
PST.CLG.980424 and  The Nuffield Foundation grant number NAL/32601,
and a grant from the Leverhulme Trust.
%
%
%%%%%%%%%%%%%%%%%%%%%%%%%%%%%%%%%%%%%%%%%%%%%%%%%%%%%%5
%

%
\end{document}